\documentclass[12pt,a4paper]{article}
\usepackage{amsmath,amsfonts,xcolor,amsthm,multirow,algorithm,algorithmicx,subfig}
\usepackage{algpseudocode}
\usepackage[pdftex,colorlinks=true]{hyperref}
\definecolor{darkblue}{rgb}{0,0,.6}
\hypersetup{citecolor=darkblue,linkcolor=darkblue,urlcolor=darkblue}
\usepackage[font={footnotesize,it}]{caption}
\usepackage[top=1.2in, bottom=1.2in, left=1.1in, right=1.1in]{geometry}
\usepackage[title]{appendix} 

\DeclareCaptionStyle{italic}[justification=centering]{labelfont={bf},textfont={it},labelsep=colon}
\usepackage{graphicx,psfrag,epsf}
\usepackage{enumerate}
\usepackage{natbib, orcidlink}
\usepackage{url, setspace}
\usepackage{booktabs, subfig, bm, paralist,mathpazo,tikz,todonotes,longtable,microtype,dsfont,rotating}
\newtheorem{theorem}{Theorem}
\newtheorem{axiom}{Axiom}
\newtheorem{prop}{Proposition}
\newtheorem{example}{Example}%
\newtheorem{definition}{Definition}%

\newcommand{\blind}{0}

\newcommand{\Rlogo}{\protect\includegraphics[height=1.8ex,keepaspectratio]{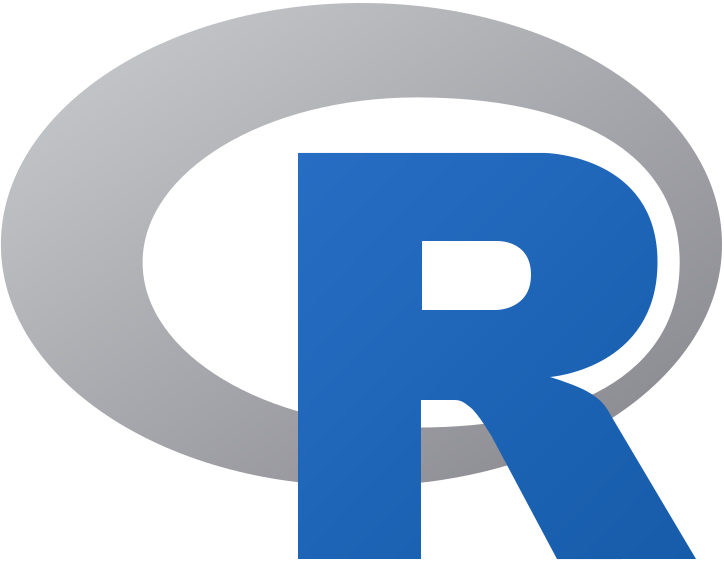}}

\graphicspath{{plots/}}

\newsavebox\CBox

\date{}

\begin{document}

\def\spacingset#1{\renewcommand{\baselinestretch}%
{#1}\small\normalsize} \spacingset{1}

\if0\blind
{
  \title{\bf Mortality Models Ensemble via Shapley Value}
  \author{\normalsize Giovanna Bimonte \quad Maria Russolillo\\
  \normalsize Department of Economics and Statistics \\ \normalsize University of Salerno \\
  \\
  \normalsize    Han Lin Shang \\
  \normalsize   Department of Actuarial Studies and Business Analytics\\ \normalsize Macquarie University\\
\\
  \normalsize    Yang Yang \\
  \normalsize   School of Information and Physical Sciences\\ \normalsize University of Newcastle
  }
  \maketitle
} \fi

\if1\blind
{
    \title{\bf Mortality Models Ensemble via Shapley Value}
  \maketitle
} \fi

\begin{abstract}
Model averaging techniques in the actuarial literature aim to forecast future longevity appropriately by combining forecasts derived from various models. This approach often yields more accurate predictions than those generated by a single model. The key to enhancing forecast accuracy through model averaging lies in identifying the optimal weights from a finite sample. Utilizing sub-optimal weights in computations may adversely impact the accuracy of the model-averaged longevity forecasts. By proposing a game-theoretic approach employing Shapley values for weight selection, our study clarifies the distinct impact of each model on the collective predictive outcome. This analysis not only delineates the importance of each model in decision-making processes, but also provides insight into their contribution to the overall predictive performance of the ensemble.

\vspace{.1in}
\end{abstract}

\newpage
\spacingset{1.5}

\section{Introduction}\label{sec1}

Accurate mortality forecasting plays a fundamental role in the actuarial industry, as it is essential for designing and pricing life insurance products. The dynamic nature of mortality necessitates that actuaries utilize appropriate tools for forecasting future longevity. While some mortality models have gained significant popularity in recent times [see, e.g., Lee \& Carter (\citeyear{LC92})], over-reliance on a limited set of models can exacerbate the adverse effects of model misrepresentation, parameter uncertainty, and overfitting. Choosing an inappropriate model to generate forecasts employed in measuring the longevity risk of a population will inevitably lead to the subsequent inaccurate pricing of life insurance products. To address this issue, we explore a model ensemble approach to increase the likelihood of selecting suitable mortality models. To this aim, we consider several families of mortality models, each with its own characteristics, strengths and weaknesses. To exploit the strengths of these models, we propose a model ensemble approach where the concept of Shapley value emerges as a vital tool for assigning weights, providing insights into the relative importance of each model within the ensemble. In this study, our proposal involves the development of an additive feature attribution approach, aiming at explaining ensemble predictions through the computation of the marginal contribution of individual models. This clarifies the individual impact of each model and provides a deeper understanding of their contributions to the overall predictive accuracy of the ensemble. This innovative approach assures more accurate mortality projections and a more sophisticated comprehension of the role that each model plays in the decision-making process.

Combining predictions from various mortality models can produce more reliable point and interval forecasts than any specific model [see, e.g., Chang \& Shi (\citeyear{CS23})]. In particular, model averaging techniques are often used to improve mortality forecast accuracy due to the simplicity of the methods and good performance  [see, e.g., Shang \citeyear{Shang12}]. Unlike the model averaging approach, our proposed method aims to optimize mortality forecast accuracy by assigning weights to models according to the Shapley value from cooperative game theory. To the best of our knowledge, this is the first time that a game-theoretic approach employing Shapley values for weight selection is used in a mortality model ensemble. A major contribution of this paper is showcasing multiple optimal weights in combinations consistently yielding improved predictions.

The approach proposed in this paper falls in the first of the two primary categories of model averaging methods, specifically, choosing optimal weights or selecting superior models. When choosing the optimal weights, it is typical to use forecasts generated by all candidate models as inputs to determine the suitable combinations. Bates \& Granger (\citeyear{BG+69}) introduced a technique to combine various individual forecasts into a consolidated set to effectively minimize forecasting errors. In more recent literature, Shang (\citeyear{Shang12}) presented the most comprehensive comparative analysis to date on combining predictions from diverse mortality models. The author demonstrated that utilizing weights based on population size contributes to generating more precise point forecasts of age-specific life expectancies than any individual model. To improve forecasting performances of stochastic mortality models, Coppola et al. (\citeyear{CRS+19}) proposed assigning weights to each selected model based on the Akaike Information Criterion (AIC). Recently, Barigou et al. (\citeyear{BGL+23}) presented two methods based on leave-future-out validation and compared them to the standard Bayesian Model Averaging (BMA) based on marginal likelihood. They demonstrated that both methods are more robust to model misspecifications and have better prediction performances than the standard BMA approach when evaluated on an out-of-sample forecasting criterion.

Instead of employing forecasts directly from different candidate models as inputs, an alternative set of methods for choosing optimal weights in model averaging involves considering model ensembles. Wolpert (\citeyear{Wolpert92}) proposed stacked generalisation ensembles that combine candidate models into a prediction loss function via a learning process. Breiman (\citeyear{Breiman04}) proposed stacked regressions that form linear combinations of candidate models as predictors to improve prediction accuracy. The model ensembles are particularly useful when the true data-generating model is not on the list of candidate models. Kessy et al. (\citeyear{KSV+22}) recently demonstrated that stacked regression ensembles improve mortality forecasting accuracy. Our proposed method follows this line of study and builds mortality model ensembles according to Shapley values of candidate models.

As regards the model averaging methods that select superior models, they are conceptually different from our proposed method. Samuels \& Sekkel (\citeyear{SS17}) put forward a novel method of trimming candidate models before combination based on the model confidence set (Hansen et al. \citeyear{HLN11}). The authors selected a set of superior models conditional on their past out-of-sample forecasting performances. Shang \& Haberman~(\citeyear{SH18}) applied the model confidence set to Japanese age-specific mortality rates at national and sub-national levels. Their proposed method selects a set of superior models from a collection of mortality models based on their in-sample forecast errors. Equal weightings are then applied to the selected models to facilitate forecast averaging.

The paper is organized as follows. In Section~\ref{sec:2}, we discuss major time series extrapolation models, which will be used to test the proposed model ensemble strategy. In Section~\ref{sec:3}, we discuss the main characteristics of model ensembles via machine learning, as well as the Shapley value properties compared to other weighting schemes. Section~\ref{sec:4} is dedicated to the empirical analysis, where we train different mortality models on historical data and combine the results into an ensemble using Shapley values. Section~\ref{sec:5} provides concluding remarks.

\section{Time series extrapolation models}\label{sec:2}

Table~\ref{tb1} lists a compilation of popular extrapolative mortality forecasting methods used by statisticians and actuaries in recent decades [see, e.g., Shang (\citeyear{SH18}) and Shang et al. (\citeyear{SBH11})]. These models will serve as the basis for demonstrating the proposed model ensemble strategy.  We are aware that many other models could have been considered in our analysis. Nevertheless, our choice was influenced by an earlier study conducted by Shang and Haberman (2018).
The mortality models in Table~\ref{tb1} can be categorized into four groups based on their structural compositions. The Renshaw-Haberman family (models M1-M3) and the Cairns-Blake-Dowd family (models M4-M8) methods are often used in the actuarial context for pricing life annuity and pension products (Pitacco et al. \citeyear{pitacco2009modelling}). All methods in these two families are implemented in \Rlogo \ using the StMoMo package by Villegas et al. (\citeyear{andres2018stmomo}). The Lee-Carter model and its variants (models M9-M12) are widely used in the demographic field for population projections [see, e.g., Plat \citeyear{plat2009stochastic}]. The Functional Time Series models (models M13-M15) use nonparametric smoothing and functional data analysis techniques to produce accurate mortality forecasts that are robust to outlying observations (Hyndman \& Ullah \citeyear{hyndman2007robust}). The demography package by Hyndman (\citeyear{Hyndman2012a}) facilitates the implementation of models belonging to both the Lee-Carter and Functional Time Series families.

\begin{table}[!htb]
\centering
\tabcolsep 0.12in
\caption{The list of 15 mortality forecasting models considered}\label{tb1}
\begin{tabular}{@{}lll@{}}
\toprule
Family of models & Label & Model \\
\midrule
Renshaw-Haberman & M1 & Lee-Carter model with Poisson error structure \\
				& M2 & Renshaw-Haberman model \\
				& M3 & Age-period-cohort model \\
\\
Cairns-Blake-Dowd	& M4 & Cairns-Blake-Dowd model \\
				& M5 & M6 model \\
				& M6 & M7 model \\
				& M7 & M8 model \\
				& M8 & Plat model \\
\\
Lee-Carter 		& M9 & Lee-Carter model with Gaussian error structure \\
				& M10 & Booth-Maindonald-Smith model \\
				& M11 & Lee-Carter model with adjustment of principal component \\ 
    & &  scores, where the method is based on life expectancy \\
				& M12 & Lee-Carter model with no adjustment to the score \\
\\
Functional Time Series & M13 & Functional time series model \\
				   & M14 & Robust functional time series model \\
				   & M15 & Product-ratio model \\								
\bottomrule
\end{tabular}
\end{table}

The models mentioned above are widely used in the analysis and projection of mortality rates. 
Every model family has its own characteristics, with distinct strengths and weaknesses. Specifically, models in the first three families adopt parametric structures in modeling mortality rates. Renshaw-Haberman and Cairns-Blake-Dowd models can capture different mortality patterns, even if Cairns-Blake-Dowd models focus specifically on modeling long-term mortality trends and age-specific improvements. Lee-Carter models are widely used in actuarial and demographic applications because of their simplicity and easy interpretability. Functional Time Series models assume stochastic processes of smooth functions in describing variations of age-specific mortality rates. This feature of functional time series models allows them to effectively capture the time-dependent patterns underlying smooth functions to model temporal dependencies of mortality data.

Thanks to their unique attributes, each family of models offers a different perspective on mortality rate modeling. Hence, selecting mortality models in practice is never easy. 
All the selected models are extrapolative, revealing a great degree of correlation in terms of the produced forecasts. This correlation probably stems from their nested structure and similar construction. Further insights into the selected models, particularly regarding biases and correlations, will be provided when discussing our results.
The model selection should be made depending on the characteristics of the available data and the specific research questions under consideration. To tackle the challenge of model selection, we propose a model ensemble approach that aims to leverage the strengths of various models to produce more accurate mortality predictions.

Our proposed method significantly reduces the risk of overfitting by incorporating different sources of model training information, effectively decreasing the model's inclination to adapt excessively to specific training data. Combining insights from multiple models yields more accurate and robust predictions than any individual model. Assigning weights to models to enhance prediction performance is often subjective and complex, especially when the candidate models diverge in prediction performances. A fair algorithm for choosing the model averaging weights requires an initial determination of each model's contribution to the ensemble. In the following section, we will explain our proposed approach for constructing model ensembles.

\section{Model ensembles via machine learning}\label{sec:3}

Prediction accuracy is critical to mortality models. Maximum accuracy requires complex models to deal with large mortality datasets in practice, which creates a trade-off between accuracy and interpretability. We consider a flexible and adaptive approach to deal with the complexity of mortality dynamics, generating highly interpretable forecasts with improved prediction accuracy. In particular, we train different mortality models on historical data and combine the results into an ensemble using Shapley values (Lundberg \& Lee \citeyear{lundberg2017}). The model ensemble can be regarded as an interpretable approximation of the original models specified by the following definition. 

\begin{definition}
    The model space $\mathcal{F}$ is the Cartesian product of $n$ models, $N = \{1,2, \ldots,n\}$, $\mathcal{F} = F_1\times F_2 \times \ldots \times F_n$, where each model $F_i$ is a finite set of model values.
\end{definition}

\begin{definition}
    A classifier $f$ is a mapping from the model space $\mathcal{F}$ to a normalized
 $|N|$-dimensional
space,  $f : \mathcal{F} \rightarrow [0,1]^{|N|}$, where $N$ is a finite set.
\end{definition}

We define a classifier function to assign to each candidate model a normalized weight representing the model's contribution to forecasts after model averaging. As detailed in Section~\ref{sec:4}, we divide the whole dataset into training, validation and testing sets and use the training set to build the classifier function. We define a cooperative game between models in characteristic function form and construct the classier function with the help of a well-known concept in cooperative game theory, namely the Shapley value [see, e.g., Roth (\citeyear{roth1988shapley})]. 

\subsection{Coalitional game approach}

In this paper, we use models as players in a cooperative game, and, through the Shapley value, determine the weight assigned to each model within the assembled one. The Shapley value allows us to analyse the individual contribution of the models to the forecasts. The purpose is to build an assembled model for the forecast that considers the influence or importance of the individual models.

Let the set $N$ comprise all possible candidate models and is called the grand coalition.
 
\begin{definition}
A coalition game is a tuple $(N, \nu)$, where $N$ is a finite set of n players, and $\nu: 2^N \rightarrow \Re$ is a characteristic function such that $\nu(\emptyset)=0$.
\end{definition}

\noindent
Each coalition represents a subset of models. Value division in coalition formation is usually studied in characteristic function games, where each potential coalition $S$ has a value $\nu(S)$ that it can obtain. The characteristic function $\nu(.)$ denotes the worth of each coalition, where $\nu(N)$ represents the worth of the grand coalition.
\begin{example}\label{ex1}
Consider a cooperative game with three agents, defined by the characteristic function:
$\nu(\emptyset) = 0$, $\nu(\{1\}) = 1$, $\nu(\{2\}) = 2$, $\nu(\{3\}) = 3$, $\nu(i, j) = 4$, $\nu(N) = 8$.
\end{example}

A solution concept is a vector $\phi(\nu)=(\phi_1,\ldots, \phi_n) \in \Re^N$  that represents the allocation of the total worth to each player. A solution concept for a coalitional game in characteristic function can be featured by some desirable properties described by the following axioms.

\begin{axiom}\label{1}
    A solution is efficient if $\sum\limits_{i=1}^{n}\phi_i(\nu)= \nu(N)$.
\end{axiom}

The principle of efficiency implies that the coalition game fairly attributes each model's contribution to the final output. This means that weights assigned to candidate models accurately reflect their contributions to the combined forecasts in the context of model averaging. Efficiency also implies that the collective contribution of the models to the final output accurately reflects the information gathered by all models in the training process.

The efficiency axiom outlines an ideal situation where the combination of models is conducted to accurately account for each model's predictive power to the solution. However, highly correlated models could hinder model averaging efficiency and stability. This is because models with a high correlation may offer overlapping information, leading to less unique weighting for each model.

\begin{axiom}\label{2}
A solution is symmetric if, for every $ S \subseteq N$ and $i,j \notin S$, $\nu(S \cup \{i\}) = \nu(S \cup \{j\})$, then $\phi_i(\nu) = \phi_j(\nu)$.  
\end{axiom}

Symmetry in model averaging implies that similar performing models receive comparable weights within the averaging process. However, in the presence of highly correlated models, symmetry might result in excessively distributed weights, overlooking subtle performance differences among models.

\begin{axiom}\label{3}
    The agent $i$ is a dummy player if  $\nu(S \cup \{i\}) = \nu(S)$, with $ S \subseteq N$ and $i \notin S$,  then $\phi_i(\nu) =0$. 
\end{axiom}

The `dummy player' axiom focuses on models within the ensemble that may not significantly impact the final prediction in the context of model averaging. When a model is extremely similar to other candidates in predictive capabilities, its inclusion or exclusion does not significantly change the output forecasts. Such models whose impact on the predictive power of the ensemble is negligible are considered redundant in model averaging. Detecting and excluding these minimal-impact models in our proposed framework helps simplify the ensemble. Therefore, we improve the computational efficiency of the proposed algorithm without significantly affecting the ensemble's predictive power.

\begin{axiom}\label{4}
    A solution is additive if for any pair of games $\nu,w$ we have  $\phi(\nu+w)= \phi(\nu)+\phi(w)$, where $(\nu+w)(S) = \nu(S)+w(S)$  for all $ S \subseteq N$.
\end{axiom}

The additivity axiom suggests that the weighted sum of candidate models' predictions forms the output predictions of the ensemble. It implies that the predictive performance of each selected model influences the ensemble's overall predictive performance in the model averaging process. When models are highly correlated, the additivity axiom might be undermined since these models could provide redundant information. Due to this reason, the unique contribution of each model may not be accurately reflected in the combined prediction. Therefore, highly correlated models could compromise the assignment of weights,
leading to a biased model ensemble and inferior forecasts. To utilize these axioms and construct a model ensemble capable of producing accurate forecasts, we use machine learning techniques to compute Shapley values.

\subsection{Shapley values via machine learning}

Driven by the principle that merging multiple predictive models leads to better results, we build ensembles of models to consolidate information. This ensemble strategy creates a pool of predictive models combined in various ways. To attribute the relative importance of each model within the ensemble, we consider the Shapley value defined in Theorem~\ref{thm1}. As a critical evaluation tool, the Shapley value not only highlights the significance of individual models in decision-making but also aids in interpreting their respective impact on the overall predictive capability of the ensemble.

\begin{theorem}[Shapley \citeyear{Shapley1953}]\label{thm1}
For a game $(N, \nu)$ there exists a unique solution $\phi$ which satisfies axioms \ref{1} to \ref{4} and it is the Shapley value:
\[
\phi_i(\nu) = \sum\limits_{S \subseteq N\setminus \{i\}} \frac{(n- |S|-1)! |S|!}{n!} \left(\nu(S \cup \{i\})-\nu(S) \right), \quad i=1,\ldots,n.
\]
\end{theorem}

\noindent
For example~\ref{ex1}, we can easily compute the Shapley values for each player:
\begin{align*}
     \phi_1(v) &= \frac{1}{6}[(1 - 0) + (4 - 0) + (2 - 0) + (4 - 3)] + \frac{1}{6}(8 - 0) = \frac{8}{3}\\
   \phi_2(v) &= \frac{1}{6}[(2 - 0) + (2 - 0) + (4 - 0) + (4 - 3)] + \frac{1}{6}(8 - 0) = \frac{17}{6}\\
   \phi_3(v) &= \frac{1}{6}[(3 - 0) + (4 - 0) + (4 - 0) + (4 - 3)] + \frac{1}{6}(8 - 0) = \frac{10}{3}\\
\end{align*}

The Shapley value has been widely adapted for interpreting machine learning models. The concept of the Shapley value in explaining predictive models is to decompose a targeted prediction value into contributions of features [see, e.g., \v{S}trumbelj and Kononenko (\citeyear{SK2014}) and Lundberg and Lee (\citeyear{lundberg2017})]. Here, the prediction takes the place of the payoff, and the features stand in for the players under the general framework of game theory. For instance, \v{S}trumbelj \& Kononenko (\citeyear{SK2010}) presented a general method involving Shapley values for individual predictions of classification models. Inspired by Aas et al. (\citeyear{AJL21}), we consider a machine learning algorithm to train the predictive models. Specifically, we regard individual forecasts produced by the mortality models described in Table~\ref{tb1} as $N$ features and attempt to use them to approximate the actual observations outside the training set. This process represents a typical scenario in supervised learning.

Let $m_{x,g,t}$ denote the actual mortality rates for gender $g = \{\text{female}, \text{male}\}$ observed at ages $x = x_1, x_2 , \ldots, x_n$ in year $t = 1, \ldots, T$ and $\widehat{m}_{x,g,\mathcal{T}+h|\mathcal{T}}^{(i)}$ denote the $h$-step-ahead forecast for $m_{x,g,t}$ obtained by training the $i\textsuperscript{th}$ ($i = 1, \ldots, N$) forecasting model $F^{(i)}$ on an arbitrary mortality sample $\mathcal{T}$.

We compute Shapley values to measure each predictive model's contribution to a coalitional forecast $\widehat{m}_{x,g,\mathcal{T}+h|\mathcal{T}}$ that adopts the following decomposition: 
\begin{equation}\label{eq1}
\widehat{m}_{x,g,\mathcal{T}+h|\mathcal{T}}= \mathbf{E}(\widehat{m}_{x,g,\mathcal{T}+h|\mathcal{T}})+\sum\limits_{i=1}^{N}\phi_i.
\end{equation}

The Shapley values $(\phi_1, \phi_2, \ldots, \phi_N)$ explain the difference between the ensemble prediction $\widehat{m}_{x,g,\mathcal{T}+h|\mathcal{T}}$ and the mean of all considered mortality process at time $\mathcal{T}+h$. 
Equation (\ref{eq1}) presents the collective mortality prediction \(\widehat{m}_{x,g,\mathcal{T}+h|\mathcal{T}}\) for a specific instance \(x\), gender \(g\), and time period \(\mathcal{T}+h\) in the context of mortality models. The overall prediction is obtained by summing the expected prediction (\(\mathbf{E}(\widehat{m}_{x,g,\mathcal{T}+h|\mathcal{T}})\)), representing the average prediction across all considered models, and the sum of Shapley values (\(\sum_{i=1}^{N}\phi_i\)), indicating the contribution of each model to the difference between the collective prediction and the mean of all considered mortality processes at time \(\mathcal{T}+h\). The Shapley values (\(\phi_1, \phi_2, \ldots, \phi_N\)) clarify how each model's prediction contributes to this difference, providing insights into the impact of individual models on the ensemble prediction relative to the overall mean prediction. A model of this form employing only additive feature assignment enjoys the properties outlined in axioms~\ref{1} to~\ref{4}. In this study, we use an additive feature attribution method to explain ensemble predictions by calculating the marginal contribution of each model. Consider $F^{(1)}, F^{(2)}, \dots, F^{(N)}$ be the $N$ mortality forecasting models. Let $(\omega_1, \omega_2, \dots, \omega_N)$ be the corresponding weights assigned to each forecast. The combined multiple forecasting models will be
\[
F^{(c)} = \omega_1 F^{(1)} + \omega_2 F^{(2)} + \cdots + \omega_m F^{(N)}, 
\]
where $(\omega_1, \dots, \omega_N)\geq 0$ and $\sum_{i=1}^N \omega_i = 1$.

In Section \ref{Selection}, we describe the procedures used to determine optimal weights via Shapley value, including other approaches commonly employed for averaging purposes.

\subsection{Approximated Shapley Value}

Applying game theory to solve real-world problems is challenging as the exact solutions are often infeasible. In practice, applications of machine learning techniques require finding good approximate solutions efficiently within a reasonable time. The computation of the Shapley value needs to evaluate an exponential number of characteristic functions, which is extremely time-consuming (Dumitrescu et al. \citeyear{dumitrescu2022machine}). Due to the exponential time complexity of Shapley values computation, we adopt a sampling approximation algorithm to exclude some subsets of models (\v{S}trumbelj and Kononenko \citeyear{SK2010}). We show that this approximated algorithm yields a Shapley value with equivalent properties.

We first introduce an alternative definition of the Shapley value before showing the approximation method. Consider $O: \lbrace1, \ldots, n\rbrace \rightarrow \lbrace1, \ldots, n\rbrace$ as a permutation where a player is assigned to the position $k$, represented as $O(k)$. Let $\pi(N)$ be the set of all permutations of the $N$ players. Given a permutation $O$, we denote by $\text{Pre}^i(O)$  the set of predecessors of the
player $i$ in the permutation $O$, that is, $\text{Pre}^i(O)= \lbrace O(1),\ldots, O(k-1)\rbrace$, if $i=O(k)$. The Shapley value can then be expressed in the following way:
\[
\phi(\nu)= \sum\limits_{O \in \pi(N)} \frac{1}{n!} \left(\nu(\text{Pre}^i(O) \cup \{i\})-\nu(\text{Pre}^i(O)) \right), \quad i=1, \ldots, n.
\]

It is easy to observe that Shapley values can be expressed as a player's expected marginal contribution when randomly added to a coalition. By randomly sampling multiple permutations and calculating the average marginal contributions made by each player, we can estimate the approximated Shapley values associated with each model. We consider Shapley value approximation methods to reduce complexity in computation while deploying machine learning algorithms to build the forecasting model ensemble. Castro et al. (\citeyear{castro2009}) first proposed approximating the Shapley value in linear time using sampling techniques. They showed that Shapley estimates were efficient if the value of each coalition can be computed in polynomial time\footnote{In our framework, we assume that all different orders have equal probability.}.

Specifically, a unique sampling process for all players $i \in \lbrace 1, \ldots, N\rbrace$ is considered in estimating the Shapley value. The sampling process $P$ is the set of all possible permutations of $N$ players, i.e. $P = \pi(N)$. The characteristics observed in each sampling unit, $O \in \pi(N)$, are the marginal contributions of players in the permutation $O$
\[
x(O)=(x(O)_1, \ldots, x(O)_N),
\]
with $x(O)_i = \nu(\text{Pre}^i(O) \cup \{i\})-\nu(\text{Pre}^i(O))$. Each iteration considers a random permutation of the set of players, and an approximated Shapley value $\widehat\phi_i$ is estimated for each player. The marginal contributions of the players in the sampled permutation are the mean of all marginal contributions and are added to the approximated Shapley values of the previous iteration. The Shapley value of a model can be represented as the expected value of the weighted marginal contribution to a randomly sampled coalition S. This coalition S is uniformly sampled from all possible coalitions rather than considering an exhaustive weighted sum. Using the Monte Carlo method, we obtain an unbiased approximation of the Shapley values. The approximated Shapley value $\widehat{\phi}_i$ is the averaged individual marginal contributions over the sample M given by
\[
\widehat\phi_i= \frac{1}{M} \sum\limits_{O \in M}x(O)_i.
\]

The vector $\widehat\phi=(\widehat\phi_1, \ldots, \widehat\phi_n)$ contains the mean of the marginal contributions of the entire population. The $\widehat\phi$ has been shown to have some desirable properties.

\begin{prop}\citeauthor{castro2009} (\citeyear{castro2009})
The estimator $\widehat\phi$ is unbiased.
\end{prop}

The estimator $\widehat\phi=(\widehat\phi_1,\ldots,\widehat\phi_n)$ is efficient as the estimated Shapley values converge to their true counterparts as the sample size increases. The estimator's efficiency depends on the computation complexity associated with each random sample. If the computation of marginal contributions can be done in polynomial time, the estimator $\widehat\phi=(\widehat\phi_1,\ldots,\widehat\phi_n)$ can provide an efficient approximation of the Shapley values (Castro et al. (\citeyear{castro2009})).

\begin{prop}
The estimator $\widehat\phi$ is symmetric (Axiom~\ref{2}).
\end{prop}
\begin{proof}
To prove that the estimator $\widehat{\bm{\phi}}=(\widehat\phi_1,\ldots,\widehat\phi_n)$ is symmetric, we need to show that the approximated Shapley values are invariant under permutations of players. In other words, if we permute the order of models, the estimated values should remain the same.

Let us consider two models, $i$ and $j$, and assume that $i$ and $j$ are interchangeable in any permutation. We will show that $\widehat{\phi}_i = \widehat{\phi}_j$. The estimated Shapley value for player $i$, denoted by $\widehat\phi_i$, is calculated as the average of the marginal contributions of player $i$ in all sampled permutations. Similarly, the estimated Shapley value for player $j$, denoted as $\widehat\phi_j$, is the average of the marginal contributions of player $j$ in the same set of sampled permutations.

Now, consider a permutation $O\in\pi(N)$ and its corresponding marginal contribution vector $x(O)=(x(O)_1, \ldots, x(O)_n)$, where $x(O)_i = \nu(\text{Pre}^i(O) \cup \{i\}) - \nu(\text{Pre}^i(O))$. If we interchange players $i$ and $j$ in the permutation $O$, a new permutation $O'$ is obtained. Under this new permutation, the marginal contributions vector becomes $x(O')=(x(O')_1, \ldots, x(O')_n)$, where $x(O')_k = x(O)_{\sigma(k)}$ for all players $k$, and $\sigma$ is the permutation that swaps $i$ and $j$.

Since the order of the players is the only difference between $O$ and $O'$, it follows that $\text{Pre}^i(O') = \text{Pre}^j(O)$ and $\text{Pre}^j(O') = \text{Pre}^i(O)$. Therefore, the marginal contributions of player $i$ in $O$ and player $j$ in $O'$ are the same, i.e., $x(O)_i = x(O')_j$. Considering all sampled permutations and their corresponding marginal contributions, we can see that for each pair of sampled permutations $O$ and $O'$, the marginal contributions of player $i$ in $O$ and player $j$ in $O'$ are equal. Consequently, the average of these marginal contributions will be the same, which gives $\widehat\phi_i$ and $\widehat\phi_j$ respectively. This symmetry property is analogous to the standard Shapley value, where the Shapley value is the same for interchangeable players.
\end{proof}

\begin{prop}
    The estimator $\widehat\phi$ is additive (Axiom~\ref{4}).
\end{prop}
\begin{proof}
To demonstrate the additive property, we need to show that the sum of the estimated Shapley values for any two models $i$ and $j$ equals the estimated Shapley value of their coalition. Let $\widehat\phi_i$ and $\widehat\phi_j$ be the estimated Shapley values for players $i$ and $j$, respectively. We want to show that $\widehat\phi_i + \widehat\phi_j = \widehat\phi_{ij}$, where $\widehat\phi_{ij}$ is the estimated Shapley value for the coalition of players $i$ and $j$.

Consider a permutation $O\in\pi(N)$ and its corresponding marginal contributions vector $x(O)=(x(O)_1, \ldots, x(O)_n)$, where $x(O)_k = \nu(\text{Pre}^k(O) \cup \{k\}) - \nu(\text{Pre}^k(O))$. The estimated Shapley value for the coalition is calculated as the average of the marginal contributions of the coalition in all sampled permutations given by
\[
\widehat\phi_{ij} = \frac{1}{M} \sum_{O \in M} x(O)_{ij}.
\]
where $x(O)_{ij} = \nu(\text{Pre}^{ij}(O) \cup \{i,j\}) - \nu(\text{Pre}^{ij}(O))$ and $Pre^{ij}(O)$ is the set of predecessors of the coalition $\{i,j\}$ in $O$. Since $\text{Pre}^{ij}(O) = \text{Pre}^i(O) \cap \text{Pre}^j(O)$, we can rewrite the marginal contributions of the coalition as:
\[
x(O)_{ij} = \nu(\text{Pre}^i(O) \cup \text{Pre}^j(O) \cup \{i,j\}) - \nu(\text{Pre}^i(O) \cup \text{Pre}^j(O)).
\]

Consider the sum of the estimated Shapley values for players $i$ and $j$:
\[
\widehat\phi_i + \widehat\phi_j = \frac{1}{M} \sum_{O \in M} x(O)_i + \frac{1}{M} \sum_{O \in M} x(O)_j.
\]
By rearranging the equation, we have:
\[
\widehat\phi_i + \widehat\phi_j = \frac{1}{M} \sum_{O \in M} (x(O)_i + x(O)_j).
\]
Note that the sum of the marginal contributions of players $i$ and $j$ in a permutation $O$ is equal to the marginal contribution of the coalition $\{i,j\}$:
\[
x(O)_i + x(O)_j = x(O)_{ij}.
\]
It can be immediately seen that
\[
\widehat\phi_i + \widehat\phi_j = \frac{1}{M} \sum_{O \in M} x(O)_{ij} = \widehat\phi_{ij}.
\]
\end{proof}

\begin{prop}
    The estimator $\widehat\phi$ enjoys the property of a dummy player (Axiom~\ref{3}).
\end{prop}

\begin{proof}
We will show that the estimated Shapley value for a player who does not contribute to any coalition is zero. Let us consider a model $i$ who is a potential dummy player. We want to show that $\widehat\phi_i = 0$ if model $i$ does not contribute to any coalition.

If player $i$ does not contribute to the value of any coalition, it means that the marginal contributions of player $i$ in all permutations $O\in\pi(N)$ are zero. In other words, $x(O)_i = 0$ for all $O\in\pi(N)$. Consider the expression for the estimated Shapley value $\widehat\phi_i$:
\[
\widehat\phi_i = \frac{1}{m} \sum_{O \in M} x(O)_i.
\]
Since $x(O)_i = 0$ for all $O\in\pi(N)$, the proof is complete.
\end{proof}

\subsection{Mean Squared Error and related axioms}

We compare the proposed Shapley value model ensemble method to a Mean Squared Error (MSE) approach that assigns weights to candidate models based on their averaged quadratic forecasting errors. Under this criterion, models with higher predictive precision are awarded higher weights. 

We define the MSE classifier for a given gender \textit{g} and time horizon \textit{h} as follows: 
\[
\text{MSE}_i(g,h) = \frac{1}{101\times (11-h)} \sum_{\zeta = 0}^{10-h} \sum_{j=1}^{101} (m_{x_j,g,\mathcal{T}_1+\zeta+h}^{(i)} - \widehat{m}_{x_j,g,\mathcal{T}_1+\zeta+h|\mathcal{T}_1+\zeta}^{(i)})^2.
\]
where $\widehat{m}_{x_j,g,\mathcal{T}_1+\zeta+h|\mathcal{T}_1+\zeta}^{(i)}$ stands for an $h$-step-ahead forecast produced by the $i\textsuperscript{th}$ predictive model based on the training sample ending in year $\mathcal{T}_1+\zeta-h$. The $x_j \in \{0,1,2,\ldots, 100\}$ and $g \in \{\text{female}, \text{male} \} $ parameters denote the age and gender of the considered population, respectively. This section also investigates if Axioms~\ref{1}-\ref{4} apply to the MSE classifier. 
\begin{prop}\label{Eff}
The $\text{MSE}_i(g,h)$ does not satisfy the Efficiency axiom, i.e., the sum of the individual model's MSE values does not equal the total MSE of the ensemble.
\end{prop}
The proof is in the Appendix~\ref{secA1}.

The MSE measures the discrepancy between a model's predictions and the actual values but does not refer to the model weighting or performance variation within a set of models.

\begin{prop}\label{Simm}
The $\text{MSE}_i(g,h)$ does not satisfy the Symmetry axiom, i.e., swapping the players' order leads to different MSE values.
\end{prop}
The proof is in the Appendix~\ref{secA1}.

\begin{prop}\label{Add}
The $\text{MSE}_i(g,h)$ does not satisfy the Additivity axiom, i.e. the sum of individual MSE values might not equal the MSE of the combined group.
\end{prop}
The proof is in the Appendix~\ref{secA1}.

In general, it is not possible to determine whether $\text{MSE}_1 + \text{MSE}_2$ equals $\text{MSE}_{c}$ without specific information about the individual MSE values and the forecast errors. The additivity of MSE depends on the specific formulation of the MSE equation and the relationship between the forecast errors of the individual groups.

\begin{prop}\label{dummy}
The $\text{MSE}_i(g,h)$ does not satisfy the Dummy Player axiom.
\end{prop}

\begin{proof}
The dummy player property requires that if a player's contribution to the coalition is zero over all possible permutations, then this player's assigned weight should be zero. However, in the case of MSE, a player's contribution (individual or group) is determined by the squared differences between their forecasts and the actual observations. Even if a player's forecasts are consistently zero or have no impact on the MSE, their squared differences may not be zero due to the presence of actual observations. As a result, their contributions to the MSE may not be zero, violating the dummy player property.
\end{proof}

\subsection{Akaike Information Criterion and related axioms}

We also consider the Akaike Information Criterion (AIC, Akaike \citeyear{akaike1974new}) as a competing method with our proposed Shapley value-based model ensemble approach. The AIC has been widely used as a model selection tool for statistical models [see, e.g., Ding et al. (\citeyear{ding2018model})]. The efficiency of the AIC classifier heavily depends on the individual predictive model under consideration. Each candidate model's AIC estimate is determined by its model fitting and complexity. The combined model's AIC estimate does not reflect individual predictive models' complexity and fit to the data. Hence, the AIC aggregation method does not satisfy the Efficiency axiom.

As shown in Table~\ref{tab:comparison}, the AIC classifier does not satisfy the additivity, symmetry, or the dummy player axiom. 
\begin{table}[!htb]
\tabcolsep 0.2in
\centering
\begin{tabular}{@{}l|cccc@{}}
\toprule
\multirow{2}{*}{\textbf{Methodology}} & \multicolumn{4}{c}{\textbf{Axioms}} \\
& \textbf{Efficiency} & \textbf{Symmetry} & \textbf{Dummy player} & \textbf{Additivity} \\ \midrule
\textbf{Shapley value} & yes & yes & yes & yes \\
\textbf{MSE} & no & no & no & no \\
\textbf{AIC} & no & no & no & no \\ \bottomrule
\end{tabular}
\caption{Axioms: Comparison between the three different methodologies}\label{tab:comparison}
\end{table}

The additivity property requires that the total AIC of a combined group should be equal to the sum of the AIC values of individual members. This property is not satisfied as the AIC value of a combined model cannot be obtained by simply summing each candidate model’s AIC estimates. 

The symmetry property requires that reordering candidate models should not change their contributions to the combined forecast. However, swapping the candidate models leads to different AIC values of the sampled permutation. 

Finally, the AIC classifier does not satisfy the dummy player axiom since the AIC estimates for individual models are not necessarily zero, even if their presence or absence has no impact on the combined model's overall AIC value. 

\section{Empirical analysis} \label{sec:4}

To assess the predictive power of the proposed model ensemble strategy, we apply the 15 forecasting mortality models listed in Table~\ref{tb1} to the Italian national mortality rates downloaded from the HMD (\citeyear{hmd}). The choice of this country is motivated by the fact that it is one of the most developed countries with the longest sample size. The study is performed on the Italian male and female population separately, for ages from 0 up to 100+ years, considered by single calendar year and by single year of age, where the class of ages above 100 years is collected in the open age group 100+. For each country, we consider 80-year observations from 1940 to 2019.

\subsection{Selection of weights}\label{Selection}

For each gender $g$, we consider $m_{x,g,t}$ and divide the mortality observations into:
\begin{itemize}
\item a 60-year initial training set $m_{\mathcal{T}_1} = \{m_{x,g,t}: x \in [0,100+]; t = 1940, \ldots, 1999\}$,
\item a 10-year validation set $m_{\mathcal{T}_2} = \{m_{x,g,t}: x \in [0,100+]; t = 2000, \ldots, 2009\}$,
\item a 10-year testing set $m_{\mathcal{T}_3} = \{m_{x,g,t}: x \in [0,100+]; t = 2010, \ldots, 2019\}$.
\end{itemize}

Let $\widehat{m}_{x,g,\mathcal{T}_1+h|\mathcal{T}_1}^{(i)}$ denote the $h$-step-ahead point forecasts produced by fitting the $i\textsuperscript{th}$ ($i=1,\ldots,N$) mortality model to the initial training set with $m_{\mathcal{T}_1}$. More specifically, for a given gender $g$ and all integer ages $x = x_1, \ldots, x_n$, we initially fit the $i\textsuperscript{th}$ mortality forecasting model listed in Table~\ref{tb1} to the training data $m_{\mathcal{T}_1}$, and produce 1-to-10-step-ahead forecasts $\widehat{m}_{x,g,\mathcal{T}_1+h|\mathcal{T}_1}^{(i)}$, with $h=1,2,\ldots,10$. In this way, we obtain 15 sets of base forecasts.

\tikzset{decorate with/.style={fill=cyan!20,draw=cyan}}
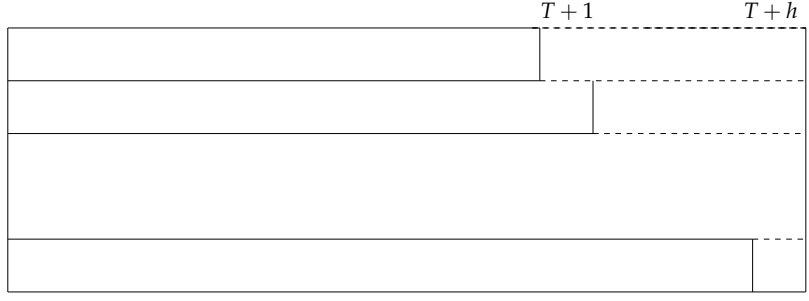
\begin{figure}[!htb]
\begin{center}
\scalebox{.7}{
\begin{tikzpicture}
\draw (0,0) -- (0, 5); 
\draw (0,0) -- (15, 0); 
\draw (10, 5) -- (10, 4);
\draw (0,4) -- (10, 4);
\draw[dashed] (10, 4) -- (15, 4);
\draw (0,3) -- (11, 3);
\draw (11, 3) -- (11, 4);
\draw[dashed] (11, 3) -- (15, 3);
\draw (0,5) -- (10, 5);
\draw[dashed] (12, 5) -- (15, 5);
\draw (0,1) -- (14, 1);
\draw (14, 1) -- (14, 0);
\draw[dashed] (14, 1) -- (15, 1);
\draw (15, 0) -- (15, 5);
\draw (9.86, 5)[dashed] node[anchor=south west]{$T+1$} -- (15, 5);
\draw (15, 5)[dashed] node[anchor=south east]{$T+h$} -- (9.86, 5);
\end{tikzpicture}}
\end{center}
\caption{A sketch of expanding-window forecasting scheme.}\label{fig:N1}
\end{figure}

In Figure~\ref{fig:N1}, through an expanding window approach, we re-estimate the parameters of the considered models using an expanded training data $m_{\mathcal{T}_1+1}$ (i.e., $m_{\mathcal{T}_1+1}= \{m_{x,g,t}: x \in [0,100+]; t = 1940, \ldots, 1999, 2000\}$ contains one additional year of observation than $m_{\mathcal{T}_1}$) and produce 1-to-9-step-ahead forecasts using the re-estimated model. The process is iterated with the sample size increased by one year in each turn until the last iteration when a training set $m_{\mathcal{T}_1+9}$ is used. This process produces ten one-step-ahead forecasts, nine two-step-ahead forecasts, and so on, up to one 10-step-ahead forecast and allows for assessing the model's stability over time.

At this point, to account for model uncertainty, we introduce the model ensemble approach. Instead of using the predictions by a specific mortality model, we combine predictions from different models to produce more reliable point and interval forecasts. 

As highlighted by Shang \& Haberman (\citeyear{SH18}),  model averaging has been used to combine the strengths of several candidate models because combined forecasts are more robust to model misspecification and are more likely to produce accurate point forecasts. 

\subsubsection{Mean normalized Shapley weights}
 
Under the expanding window setting, for a particular forecasting horizon $h$, we collect all $h$-step-ahead forecasts obtained by fitting the $i\textsuperscript{th}$ model to training data with various lengths into a vector given by
\[
\widetilde{m}_{x,g,h}^{(i)} = \left\{\widehat{m}_{x,g,\mathcal{T}_1+h|\mathcal{T}_1}^{(i)}, \widehat{m}_{x,g,(\mathcal{T}_1+1)+h|\mathcal{T}_1+1}^{(i)}, \ldots, \widehat{m}_{x,g,\mathcal{T}_1+10|\mathcal{T}_1+10-h}^{(i)}\right\}.
\]
	
We then feed the vector $\widetilde{m}_{x,g,h}^{(i)}$ as the $i\textsuperscript{th}$ feature into the Shapley value computation using the \verb|fastshap| package by Greenwell (\citeyear{Greenwell21}) in \Rlogo \ by R Core Team (\citeyear{Team23}). Specifically, we follow the fastshap vignette (Greenwell \citeyear{Greenwell21}) and build a random forest as the characteristic function $\nu$ to explain the contribution of features $\widetilde{m}_{x,g,h}^{(i)}$. The random forest model is implemented by the \verb|ranger| package by Wright \& Ziegler (\citeyear{WWP17}), which by default adopts probability forest (Malley et al. \citeyear{malley2012probability}) to determine splits during the decision tree-growing process. We denote the obtained Shapley values associated with the $i\textsuperscript{th}$ feature (i.e., $i\textsuperscript{th}$ forecasting model) by $\widehat{\phi}_i (\nu; x, g,h)$.

Next, we average the absolute age-specific Shapley estimates as
\[
\overline{\phi}_i(\nu; g) = \frac{1}{10\times n} \sum_{h=1}^{10} \sum_{j=1}^n |\widehat{\phi}_i(\nu; x_j,g,h)|, \quad i = 1, \ldots, N.
\]
We proceed to normalize the mean Shapley values across $i$ as
\[
\widetilde{\phi}_i(\nu; g) = \frac{\overline{\phi}_i(\nu; g) - \frac{1}{N}\sum_{i=1}^N\overline{\phi}_i(\nu; g)}{\text{SE}(\overline{\phi}(\nu; g))},
\]
where the $\text{SE}(\overline{\phi}(\nu; g))$ is the standard error of $\{\overline{\phi}_1(\nu; g), \ldots, \overline{\phi}_{N}(\nu; g)\}$.
 
We then determine the weights as
\[
\widehat{\omega}_{i,\text{Shapley}} = \frac{\exp(\widetilde{\phi}_i(\nu; g))}{\sum_{i=1}^{N}\exp(\widetilde{\phi}_i(\nu; g))}.
\]
 
Finally, for the $i\textsuperscript{th}$ forecasting method, we obtain a set of base forecasts $\{\widehat{m}_{x,g,\mathcal{T}_1+\mathcal{T}_2+h|\mathcal{T}_1+\mathcal{T}_2}^{(1)},\ldots, \allowbreak \widehat{m}_{x,g,\mathcal{T}_1+\mathcal{T}_2+h|\mathcal{T}_1+\mathcal{T}_2}^{(15)}\}$ by fitting the corresponding model to training data with increasing sizes. Under the expanding window setting, for any pair of $h = 1,\ldots, 10$ and $\zeta = 0,\ldots,10-h$, the combined forecast is then computed as
\begin{equation}
\widehat{m}^{(c)}_{\text{Shapley}}(x,g,\mathcal{T}_1+\mathcal{T}_2+\zeta+h|\mathcal{T}_1+\mathcal{T}_2+\zeta) = \sum_{i=1}^{15} \widehat{\omega}_{i,\text{Shapley}} \widehat{m}_{x,g,\mathcal{T}_1+\mathcal{T}_2+\zeta+h|\mathcal{T}_1+\mathcal{T}_2+\zeta}^{(i)}. \label{Shap}
\end{equation}

In this framework, we determine the weights for combining the mortality projections from different models using the Shapley value and the following algorithm to enhance the accuracy of both point and interval forecasts.

\begin{algorithm}
\caption{Shapley Value-based Forecast Combination}\label{algo2}
\begin{algorithmic}[1]
\Require Forecasting horizon $h$, training data, forecasting models
\Ensure Combined forecast using Shapley values

\State Collect $h$-step-ahead forecasts: $\widetilde{m}_{x,g,h}^{(i)}$
\State Compute Shapley values $\widehat{\phi}_i (\nu; x, g,h)$

\State Calculate Shapley value averages and normalization: $\overline{\phi}_i(\nu; g)$
    and $\widetilde{\phi}_i(\nu; g)$

\State Determine weights: $\widehat{\omega}_{i,\text{Shapley}}$

\State Base forecasts for each method: $\{\widehat{m}_{x,g,\mathcal{T}_1+\mathcal{T}_2+h|\mathcal{T}_1+\mathcal{T}_2}^{(1)}, \ldots, \widehat{m}_{x,g,\mathcal{T}_1+\mathcal{T}_2+h|\mathcal{T}_1+\mathcal{T}_2}^{(15)}\}$

\State Combined forecast: $\widehat{m}^{(c)}_{\text{Shapley}}(x,g,\mathcal{T}_1+\mathcal{T}_2+\zeta+h|\mathcal{T}_1+\mathcal{T}_2+\zeta)$
\end{algorithmic}
\end{algorithm}

\subsubsection{Alternative weighting schemes}

We consider three alternative weighting schemes to assess the performance of our model ensemble approach. These schemes are well-known and widely used in the existing literature.

\subsubsection*{Equal weights}

The most straightforward method for selecting the weights is assigning equal weights to all forecasts, equivalent to taking a simple average of the base forecasts obtained by various methods. In this context, as we consider in total $N=15$ mortality forecasting models as detailed in Table~\ref{tb1}, a weight of $1/15$ is used to scale all forecasts in computing the combination, for $ h = 1,2,\ldots,10 \ \ \mbox{and} \ \ \zeta = 0,\ldots,10-h $:
\begin{equation}
\widehat{m}^{(c)}_{\text{Average}}(x,g,\mathcal{T}_1+\mathcal{T}_2+\zeta+h|\mathcal{T}_1+\mathcal{T}_2+\zeta) = \frac{1}{15}\sum_{i=1}^{15} \widehat{m}_{x,g,\mathcal{T}_1+\mathcal{T}_2+\zeta+h|\mathcal{T}_1+\mathcal{T}_2+\zeta}^{(i)} \label{EqW}
\end{equation} 

\subsubsection*{Inverse MSE weights}

The second weighting scheme we consider to compare our model ensemble approach is based on the Mean Squared Error. For the MSE computation, we consider $h$-step-ahead forecasts obtained using the training set under the expanding window setting. For a particular forecasting horizon $h = 1,2,\ldots, 10$ and $\zeta = 0,\ldots,10-h$, we compare the obtained forecasts to the actual observations and compute MSE as
\[
\text{MSE}_i(g,h) = \frac{1}{101\times (11-h)} \sum_{\zeta = 0}^{10-h} \sum_{j=1}^{101} (m_{x_j,g,\mathcal{T}_1+\zeta+h}^{(i)} - \widehat{m}_{x_j,g,\mathcal{T}_1+\zeta+h|\mathcal{T}_1+\zeta}^{(i)})^2.
\]
 
Next, we average the $\text{MSE}_i(g,h)$ over all the considered forecasting horizons as
\[
\overline{\text{MSE}}_{i}(g) = \frac{1}{10} \sum_{h=1}^{10} \text{MSE}_i(g,h).
\]
 
The weights for combining forecasts are then obtained as
\[
\widehat{\omega}_{i, \text{MSE}} = \frac{\left[\overline{\text{MSE}}_{i}(g)\right]^{-1}}{\sum_{i=1}^{15}\left[\overline{\text{MSE}}_{i}(g)\right]^{-1}}.
\]
 
Finally, for any pair of $h = 1,\ldots, 10$ and $\zeta = 0,\ldots,10-h$, we compute the combined forecast as
\begin{equation}
\widehat{m}^{(c)}_{\text{MSE}}(x,g,\mathcal{T}_1+\mathcal{T}_2+\zeta+h|\mathcal{T}_1+\mathcal{T}_2+\zeta) = \sum_{i=1}^{15} \widehat{\omega}_{i,\text{MSE}} \widehat{m}_{x,g,\mathcal{T}_1+\mathcal{T}_2+\zeta+h|\mathcal{T}_1+\mathcal{T}_2+\zeta}^{(i)}. \label{MSE}
\end{equation}

\subsubsection*{Inverse AIC weights}

The last weighting scheme we consider for our comparison purpose is based on the Akaike Information Criterion. Under the expanding window approach, for a particular $h = 1,2,\ldots, 10$ and $\zeta = 0,\ldots, 10-h$, we compute the AIC for the fitted model according to the following formula by Johnson \& Wichern (\citeyear{johnson2002applied}):

\begin{equation*}
\text{AIC}^{(i)}_{\zeta}(g) = |R_{g,\mathcal{T}_1+\zeta}^{(i)}| \ln \left(\frac{\text{RSS}}{|R_{g,\mathcal{T}_1+\zeta}^{(i)}|}\right) + 2p,
\end{equation*}
where $R_{g,\mathcal{T}_1+\zeta}^{(i)} = \{m_{x,g,\mathcal{T}_1+\zeta} - \hat{m}_{x,g,\mathcal{T}_1+\zeta}^{(i)}; x = x_1, \ldots, x_n \}$ is the residuals across the entire training set (i.e., differences between the observations in the training set and the fitted values of the $i\textsuperscript{th}$ model) and $|R_{g,\mathcal{T}_1+\zeta}^{(i)}|$ denotes the sample size (i.e., number of points) of $R_{g,\mathcal{T}_1+\zeta}^{(i)}$; the RSS is a generic notation of the residual sum of squares computed on the residuals $R_{g,\mathcal{T}_1+\zeta}^{(i)}$, and $p$ is the number of parameters in the considered model.

We average the AIC values for all $h$, across all $\zeta$ to obtain
\begin{equation*}
\overline{\text{AIC}}_i(g) = \frac{1}{11-h}\sum_{\zeta=0}^{10-h} \text{AIC}^{(i)}_{\zeta}(g),
\end{equation*}
and then we compute the weights for the combination as
\begin{equation*}
\widehat{\omega}_{g,i,\text{AIC}} = \frac{(\overline{\text{AIC}}_{i}(g))^{-1}}{\sum_{i=1}^{15}(\overline{\text{AIC}}_{i}(g))^{-1}}.
\end{equation*}

Finally, for any pair of $h = 1,\ldots, 10$ and $\zeta = 0,\ldots,10-h$, we compute the combined forecast as
\begin{equation}\label{AIC}
\widehat{m}^{(c)}_{\text{AIC}}(x,g,\mathcal{T}_1+\mathcal{T}_2+\zeta+h|\mathcal{T}_1+\mathcal{T}_2+\zeta) = \sum_{i=1}^{15} \widehat{\omega}_{g,i,\text{AIC}} \widehat{m}_{x,g,\mathcal{T}_1+\mathcal{T}_2+\zeta+h|\mathcal{T}_1+\mathcal{T}_2+\zeta}^{(i)}
\end{equation}
 
\subsection{Point and interval forecast evaluation}

An intensive numerical study has been carried out to compare the performance of our proposed strategy to the other model ensemble approaches in terms of point and interval forecast accuracy. To further investigate the choice of mortality models selected for the ensemble model, we have also computed biases and correlations of forecasts produced by 15 individual mortality models. The computation was carried out both for the projections over the period 2000-2009 and in the test set from 2010 to 2019\footnote{The Figures for the period 2010-2019 are available on request.}.

\begin{figure}[!htb]
\includegraphics[width=5in]{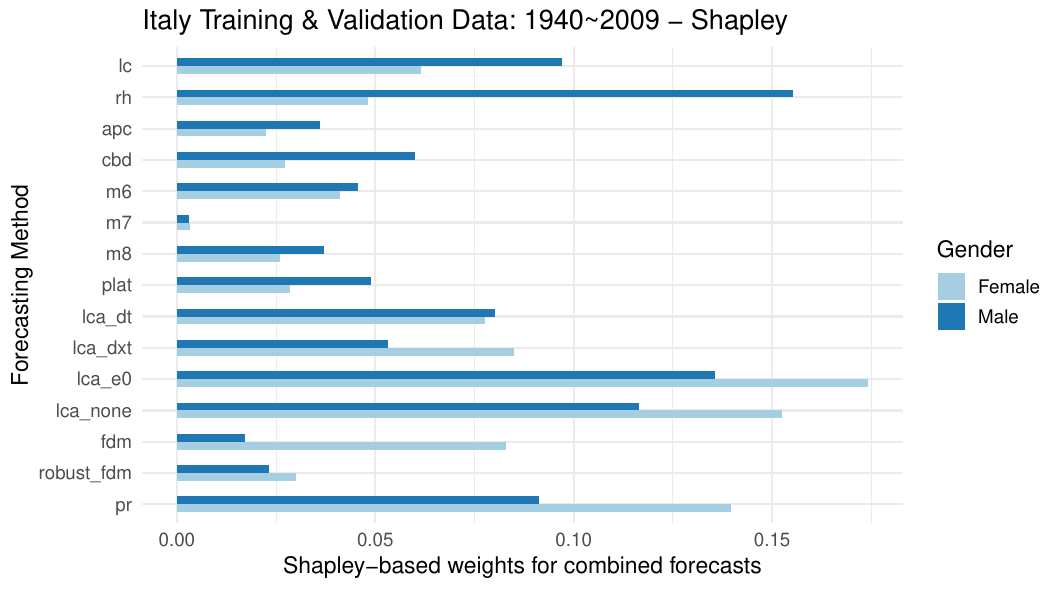}
\caption{Italy Training \& Validation Data: 1940~2009 – Shapley}\label{A1}
\end{figure}

\begin{figure}[!htb]
\includegraphics[width=4.95in]{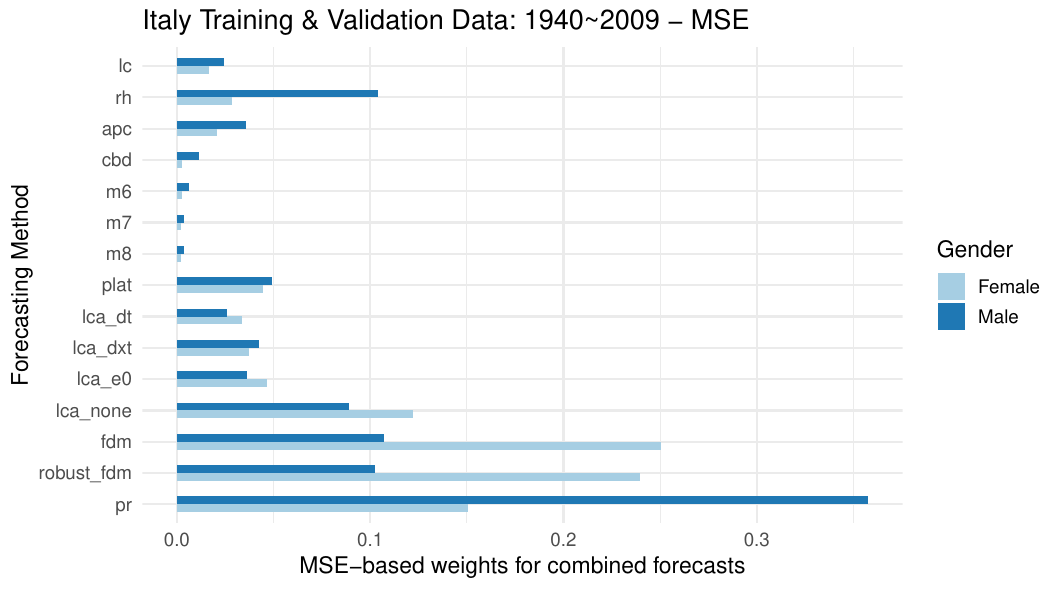}
\caption{Italy Training \& Validation Data: 1940~2009 – MSE}\label{A2}
\end{figure}

\begin{figure}[!htb]
\includegraphics[width=4.95in]{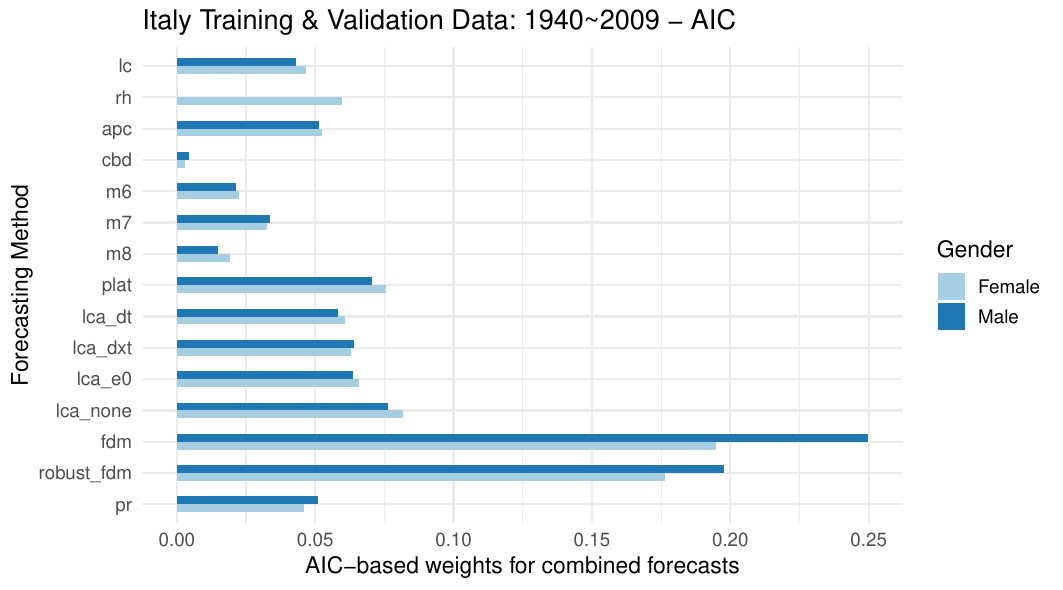}
\caption{Italy Training \& Validation Data: 1940~2009 – AIC}\label{A3}
\end{figure}

From the steps described in Section~\ref{Selection}, we produce a set of 1-to-10-step-ahead forecasts for each method. The results are 15 sets of forecasts waiting to be combined according to the weights calculated based on the four described weighting schemes. Thus, by using the validation set between 2000 and 2009, we compute the weights for each method according to~\eqref{Shap} to~\eqref{AIC}. In Figures~\ref{A1} to~\ref{A3}, we show the weights selected by the different schemes. From these Figures, we can note that the Shapley-based criterion assigns more equally distributed weights to each model in the set, denoting a greater potential for risk diversification than the other approaches.

From Figure \ref{Biases} (see Appendix \ref{secA2}), we observe a noticeable variability among the models, for the years 2000 to 2009, in terms of forecast biases. In particular, we can note that models belonging to the Renshaw-Haberman family and the Lee-Carter model with nearly all its variants (the only exception is represented by the Lee-Carter model with no adjustment to the score) are biased up. The remaining models (except the original CBD model (M4), which provides the largest overestimates for the female series) exhibit a downward bias for both male and female data. The bias, whether up or down, in the selected models underlines the necessity of applying model weighting. As aforementioned, relying just on a single model to generate forecasts, might lead to the inaccurate pricing of life insurance products. 

Figure \ref{Correlation} shows the correlations between forecasts generated by 15 individual mortality models for the years ranging from 2000 up to 2009. The figure clearly exhibits that the 15 models examined in this study generated forecasts that were highly correlated, regardless of gender. Notably, the CBD model with a cohort effect (M6) was observed to produce forecasts that were the least correlated with those of other models, particularly with M5 and M7.

Despite the strong correlation observed among the 15 models, with a few exceptions noted earlier, the Shapley value assigns widely varying weights. This suggests that the Shapley value captures additional important facets of each model. It not only underlines the importance of individual models in decision-making but also facilitates the interpretation of their respective contributions to the overall predictive capacity of the ensemble.
 
\subsubsection*{Point forecast evaluation}

To evaluate the four sets (one for each method) of the combined forecasts, we use the test set covering 10-year observations (2010–2019). Throughout the demographic package Demography (\citeyear{Hyndman2012a}) in \Rlogo, we compare the combined forecasts to the holdout samples and calculate the point and interval forecasts. Point forecast accuracy results are provided in terms of Mean Squared Error (MSE) and Mean Absolute Error (MAE) in Table~\ref{T1} and~\ref{T2}, respectively. These tables report summaries of the point forecast accuracy at the logarithm scale for Italy based on MSE and MAE criteria for one-step-ahead forecasts. Results are reported by different forecasting horizons $h = 1,\ldots, 10$ and for females and males separately for each weighting scheme.

\begin{center}
\tabcolsep 0.09in
\begin{longtable}{@{}lllllllll@{}}
\caption{Point forecast accuracy for Italy -- MSE.\label{T1}}\\
\toprule
& \multicolumn{2}{c}{$\widehat{m}^{(c)}_{\text{Average}}$} & \multicolumn{2}{c}{$\widehat{m}^{(c)}_{\text{Shapley}}$} & \multicolumn{2}{c}{$\widehat{m}^{(c)}_{\text{MSE}}$} & \multicolumn{2}{c}{$\widehat{m}^{(c)}_{\text{AIC}}$} \\\cmidrule{2-9}
$h$ & Female & Male & Female & Male & Female & Male & Female & Male \\ 
\midrule
\endfirsthead
& \multicolumn{2}{c}{$\widehat{m}^{(c)}_{\text{Average}}$} &    \multicolumn{2}{c}{$\widehat{m}^{(c)}_{\text{Shapley}}$} & \multicolumn{2}{c}{$\widehat{m}^{(c)}_{\text{MSE}}$} & \multicolumn{2}{c}{$\widehat{m}^{(c)}_{\text{AIC}}$} \\\cmidrule{2-9}
$h$ & Female & Male & Female & Male & Female & Male & Female & Male \\ 
\midrule
\endhead
\hline \multicolumn{9}{r}{{Continued on next page}} \\
\endfoot
\hline
\endlastfoot
1 & 0.0818 & 0.0806 & 0.0262 & 0.0298 & 0.0158 & 0.0121 & 0.0295 & 0.0240 \\ 
2 & 0.1130 & 0.1119 & 0.0314 & 0.0360 & 0.0174 & 0.0134 & 0.0394 & 0.0328 \\ 
3 & 0.1563 & 0.1647 & 0.0376 & 0.0484 & 0.0179 & 0.0156 & 0.0537 & 0.0499 \\ 
4 & 0.2296 & 0.2317 & 0.0510 & 0.0611 & 0.0205 & 0.0177 & 0.0841 & 0.0769 \\ 
5 & 0.3102 & 0.3345 & 0.0606 & 0.0818 & 0.0214 & 0.0205 & 0.1204 & 0.1246 \\ 
6 & 0.4118 & 0.4641 & 0.0729 & 0.1028 & 0.0222 & 0.0211 & 0.1698 & 0.1904 \\ 
7 & 0.5754 & 0.6386 & 0.0985 & 0.1382 & 0.0256 & 0.0319 & 0.2614 & 0.2910 \\ 
8 & 0.7166 & 0.7803 & 0.1151 & 0.1577 & 0.0272 & 0.0342 & 0.3387 & 0.3698 \\ 
9 & 0.8429 & 0.9372 & 0.1290 & 0.1802 & 0.0246 & 0.0366 & 0.3961 & 0.4525 \\ 
10 & 0.9649 & 1.4155 & 0.1278 & 0.3051 & 0.0231 & 0.0900 & 0.4492 & 0.7713 \\ 
\cmidrule{2-9}
Mean & 0.4402 & 0.5159 & 0.0750 & 0.1141 & 0.0216 & 0.0293 & 0.1942 & 0.2383 \\
\bottomrule
\end{longtable}
\end{center}

\begin{center}
\tabcolsep 0.09in
\begin{longtable}{@{}lllllllll@{}}
\caption{Point forecast accuracy for Italy -- MAE.\label{T2}} \\
\toprule
& \multicolumn{2}{c}{$\widehat{m}^{(c)}_{\text{Average}}$} & \multicolumn{2}{c}{$\widehat{m}^{(c)}_{\text{Shapley}}$} & \multicolumn{2}{c}{$\widehat{m}^{(c)}_{\text{MSE}}$} & \multicolumn{2}{c}{$\widehat{m}^{(c)}_{\text{AIC}}$} \\\cmidrule{2-9}
$h$ & Female & Male & Female & Male & Female & Male & Female & Male \\ 
\midrule
\endfirsthead
& \multicolumn{2}{c}{$\widehat{m}^{(c)}_{\text{Average}}$} &    \multicolumn{2}{c}{$\widehat{m}^{(c)}_{\text{Shapley}}$} & \multicolumn{2}{c}{$\widehat{m}^{(c)}_{\text{MSE}}$} & \multicolumn{2}{c}{$\widehat{m}^{(c)}_{\text{AIC}}$} \\\cmidrule{2-9}
$h$ & Female & Male & Female & Male & Female & Male & Female & Male \\ 
\midrule
\endhead
\hline \multicolumn{9}{r}{{Continued on next page}} \\
\endfoot
\hline
\endlastfoot
1 & 0.1431 & 0.1550 & 0.0967 & 0.1165 & 0.0802 & 0.0715 & 0.0977 & 0.0955 \\ 
2 & 0.1614 & 0.1766 & 0.1047 & 0.1268 & 0.0839 & 0.0775 & 0.1077 & 0.1074 \\ 
3 & 0.1865 & 0.2121 & 0.1123 & 0.1441 & 0.0851 & 0.0840 & 0.1195 & 0.1268 \\ 
4 & 0.2273 & 0.2523 & 0.1286 & 0.1623 & 0.0929 & 0.0906 & 0.1431 & 0.1513 \\ 
5 & 0.2650 & 0.3030 & 0.1370 & 0.1866 & 0.0976 & 0.1005 & 0.1624 & 0.1854 \\ 
6 & 0.3153 & 0.3634 & 0.1489 & 0.2108 & 0.0991 & 0.1027 & 0.1857 & 0.2203 \\ 
7 & 0.3887 & 0.4402 & 0.1728 & 0.2444 & 0.1050 & 0.1220 & 0.2249 & 0.2741 \\ 
8 & 0.4504 & 0.5012 & 0.1873 & 0.2681 & 0.1095 & 0.1168 & 0.2534 & 0.3048 \\ 
9 & 0.5356 & 0.5870 & 0.2073 & 0.3032 & 0.1046 & 0.1315 & 0.2902 & 0.3670 \\ 
10 & 0.6310 & 0.7318 & 0.2329 & 0.3674 & 0.0994 & 0.1842 & 0.3430 & 0.4848 \\ 
\cmidrule{2-9}
Mean & 0.3304 & 0.3723 & 0.1528 & 0.2130 & 0.0957 & 0.1081 & 0.1927 & 0.2317 \\
\bottomrule
\end{longtable}
\end{center}

\subsubsection*{Interval forecast evaluation}

In addition to point forecasts, we also compute interval forecasts to assess forecast uncertainty. Interval forecasts for all 15 considered methods are obtained following Shang \& Haberman (\citeyear{SH18}). The interval forecast accuracy is evaluated using the Mean Interval Score (Gneiting \& Raftery \citeyear{GR07}). For either gender and each year in the forecasting period, we compute a point-wise prediction interval for the forecast $\widehat{m}^{(c)}(x,g,\mathcal{T}_1+\mathcal{T}_2+\zeta+h|\mathcal{T}_1+\mathcal{T}_2+\zeta)$ at the $(1-\alpha)\times 100\%$ nominal coverage probability. For simplicity of notations, we suppress the parameters for gender and training set, i.e., we use $\widehat{m}^{(c),\text{lb}}(x)$ and $\widehat{m}^{(c),\text{ub}}(x)$ to denote the lower and upper bounds for the considered point forecast, respectively; we use $m_x$ to denote the observed mortality. A scoring rule for the interval forecast at $x$ is given by
\begin{align*}
S_{\alpha}\left[\widehat{m}^{(c),\text{lb}}(x), \widehat{m}^{(c),\text{ub}}(x), m_{x}\right] &  =  \left[\widehat{m}^{(c),\text{ub}}(x) - \widehat{m}^{(c),\text{lb}}(x)\right] \\
& + \frac{2}{\alpha}\left[\widehat{m}^{(c),\text{lb}}(x) - m_{x}\right]\mathds{1}\left\{m_{x}<\widehat{m}^{(c),\text{lb}}(x)\right\} \\
& + \frac{2}{\alpha}\left[m_{x} - \widehat{m}^{(c),\text{ub}}(x)\right]\mathds{1}\left\{m_{x}>\widehat{m}^{(c),\text{ub}}(x)\right\},
\end{align*}
where $\mathds{1}\{\cdot\}$ represents the binary indicator function, and $\alpha$ denotes a level of significance, fixed at the typically $80\%$ nominal coverage probability. In Table~\ref{T3}, forecast accuracy results are provided in terms of Mean Interval Score over the number of observations in the forecasting period. Notice that although for point forecasts, the combined projections based on the MSE criterion perform slightly better than those based on the Shapley, this is not always true as for interval forecasts the situation is reversed. As one may expect, the MSE is performing better than other approaches in the test set since the criterion by which an approach is said to perform better or worse is the same as the one used for the construction of MSE estimates. Nevertheless, as far as forecasting is usually applied to unknown dynamics, the satisfaction of properties, as in the Shapley value approach, is to be considered worthwhile when defending the reliability of projections. In other words, the innovative aspect of our approach lies in its ability to facilitate forecasting in future time periods, particularly in situations where data is unavailable or the dynamics are unknown.

\begin{center}
\tabcolsep 0.09in
\begin{longtable}{@{}lllllllll@{}}
\caption{Mean Interval Score ($\times 100$) for Italy.\label{T3}} \\
\toprule
& \multicolumn{2}{c}{$\widehat{m}^{(c)}_{\text{Average}}$} & \multicolumn{2}{c}{$\widehat{m}^{(c)}_{\text{Shapley}}$} & \multicolumn{2}{c}{$\widehat{m}^{(c)}_{\text{MSE}}$} & \multicolumn{2}{c}{$\widehat{m}^{(c)}_{\text{AIC}}$} \\\cmidrule{2-9}
$h$ & Female & Male & Female & Male & Female & Male & Female & Male \\ 
\midrule
\endfirsthead
& \multicolumn{2}{c}{$\widehat{m}^{(c)}_{\text{Average}}$} &    \multicolumn{2}{c}{$\widehat{m}^{(c)}_{\text{Shapley}}$} & \multicolumn{2}{c}{$\widehat{m}^{(c)}_{\text{MSE}}$} & \multicolumn{2}{c}{$\widehat{m}^{(c)}_{\text{AIC}}$} \\\cmidrule{2-9}
$h$ & Female & Male & Female & Male & Female & Male & Female & Male \\ 
\midrule
\endhead
\hline \multicolumn{9}{r}{{Continued on next page}} \\
\endfoot
\hline
\endlastfoot
1 & 1.9368 & 2.4307 & 1.3265 & 2.3688 & 1.7760 & 2.9342 & 1.8984 & 2.5709 \\ 
2 & 1.9909 & 2.4236 & 1.3490 & 2.1660 & 1.5325 & 2.4883 & 1.7635 & 2.2015 \\ 
3 & 2.0161 & 2.4686 & 1.3792 & 2.0222 & 1.3996 & 2.2080 & 1.6599 & 2.0256 \\ 
4 & 2.0197 & 2.4926 & 1.3077 & 1.9515 & 1.3387 & 1.9947 & 1.6447 & 1.9020 \\ 
5 & 1.9883 & 2.5096 & 1.2560 & 1.8334 & 1.1121 & 1.9159 & 1.3130 & 1.8267 \\ 
6 & 2.0172 & 2.5411 & 1.2915 & 1.8427 & 1.1002 & 1.9603 & 1.2515 & 1.8365 \\ 
7 & 2.0485 & 2.5703 & 1.3260 & 1.8355 & 1.1374 & 2.0282 & 1.2737 & 1.8601 \\ 
8 & 2.0646 & 2.5939 & 1.3551 & 1.8489 & 1.1782 & 2.0930 & 1.3070 & 1.9098 \\ 
9 & 2.0831 & 2.6021 & 1.3865 & 1.8589 & 1.2191 & 2.1256 & 1.3121 & 1.8997 \\ 
10 & 2.1138 & 2.6883 & 1.4281 & 1.9041 & 1.2696 & 2.2849 & 1.3453 & 2.0919 \\ 
\cmidrule{2-9}
Mean & 2.0279 & 2.5321 & 1.3406 & 1.9632 & 1.3064 & 2.2033 & 1.4769 & 2.0125 \\ 
\bottomrule
\end{longtable}
\end{center}

\section{Conclusion}\label{sec:5}

Model averaging techniques are particularly important in the actuarial literature for appropriately forecasting future longevity. To overcome model misspecification error, parameter uncertainty, and overfitting due to an over-reliance on a limited set of models, we propose an ensemble model approach that significantly increases the likelihood of selecting suitable mortality models. Most importantly, our proposed approach seeks to enhance the accuracy of mortality forecasts by allocating weights to models based on the Shapley value derived from cooperative game theory. The ensemble strategy involves the integration of diverse algorithms from Machine Learning, creating a pool of varied predictors. Within this framework, the Shapley value concept emerges as a crucial assessment tool for determining the relative significance of each model within the ensemble. This approach clarifies the individual impact of each model on the combined predictive model. Consequently, this evaluation not only discerns the importance of each model in the decision-making process but also facilitates an understanding of how each model contributes to the overall predictive prowess of the ensemble. To assess the predictive power of the proposed model ensemble strategy, we apply 15 well-known forecasting mortality models to the Italian national mortality rates. The proposed ensemble method involving Shapley values is verified to produce forecasts as accurate as those generated by MSE, more accurate forecasts than the equally weighted and the AIC methods, and the most precise interval forecasts. Moreover, the Shapley-based methodology not only provides better diversification of the model risk, but most importantly, respect to the other weighted method facilitates a more comprehensive interpretation of outcomes attributable to normative properties such as \textit{efficiency} and \textit{additivity}. 

To conclude, there are at least two possibilities for extending the findings of the current study:
\begin{inparaenum}
\item[1)] It would be interesting to consider more countries and run a comparative analysis to understand the best model ensemble approach for the different countries.
\item[2)] One might consider using the Shapley value for the selection of superior models.
\end{inparaenum}

\subsection*{Declarations} 

\noindent
\textbf{Conflict of interest} 
All authors declare that they have no potential conflicts of interest.

\noindent
\textbf{Funding} Not applicable 

\noindent
\textbf{Human and animal rights} This article does not contain any studies with animals performed by any of the authors. 

\newpage
\begin{appendices}
\section{}\label{secA1}
Let us consider two models with their respective MSE values denoted as $\text{MSE}_1$ and $\text{MSE}_2$, we prove the results of propositions~\ref{Eff},~\ref{Simm},~\ref{Add}.

\proof{\textit{Proposition~\ref{Eff}}}

Consider the  combined MSE, $\text{MSE}^c$, obtained by linearly combining the MSE's for both models:
\begin{align*}
\text{MSE}^c = \frac{1}{101 \times (T-h)} \sum_{\zeta = 0}^{10-h} \sum_{j=1}^{101} \left(m_{x_j,g,\mathcal{T}_1+\zeta}^{(1)} - \widehat{m}_{x_j,g,\mathcal{T}1+\zeta|\mathcal{T}_1+\zeta-h}^{(1)}\right)^2 + \\
\frac{1}{101 \times (T-h)} \sum_{\zeta = 0}^{10-h} \sum_{j=1}^{101} \left(m_{x_j,g,\mathcal{T}_1+\zeta}^{(2)} - \widehat{m}_{x_j,g,\mathcal{T}_1+\zeta|\mathcal{T}_1+\zeta-h}^{(2)}\right)^2
\end{align*}

Making a few simple algebraic steps, we obtain:
\begin{align*}
\text{MSE}^c - (\text{MSE}_1 + \text{MSE}_2) = &
\frac{1}{101 \times (T-h)} \sum_{\zeta = 0}^{10-h} \sum_{j=1}^{101} \Big[2(m_{x_j,g,\mathcal{T}_1+\zeta}^{(1)} - m_{x_j,g,\mathcal{T}_1+\zeta}^{(2)}) \\
& (\widehat{m}_{x_j,g,\mathcal{T}_1+\zeta|\mathcal{T}_1+\zeta-h}^{(1)} - \widehat{m}_{x_j,g,\mathcal{T}_1+\zeta|\mathcal{T}_1+\zeta-h}^{(2)})\Big]
\end{align*}
that is not always equal to zero, i.e. we can conclude that the MSE does not satisfy the Efficiency axiom of the Shapley value, as the sum of the individual MSE values does not equal the combined MSE.

\proof{\textit{Proposition~\ref{Simm}}}

Consider the case of reversing the order of the models in the $\text{MSE}$ formula, and we obtain a $\text{MSE}_2'$ for the first model and $\text{MSE}_1'$ for the second model. By comparing $\text{MSE}_1$ and $\text{MSE}_2'$ as well as $\text{MSE}_2$ and $\text{MSE}_1'$ , we can see that they are not equal, i.e $\text{MSE}_1 - \text{MSE}_2'  \neq 0$ and $\text{MSE}_2 - \text{MSE}_1' \neq 0$ 

This indicates that the MSE is not symmetric, as swapping the order of the players leads to different MSE values. Therefore, the MSE does not satisfy the Symmetry axiom of the Shapley value.

\proof{\textit{Proposition~\ref{Add}}}

Let us consider two groups of models, Group 1 and Group 2, with their respective MSE values, denoted as $\text{MSE}_1$ and $\text{MSE}_2$. We also have the combined group, denoted as Group 1 + Group 2, with its MSE value denoted as $\text{MSE}_{c}$.

To check the additivity, we need to compare $\text{MSE}_1 + \text{MSE}_2$ and $\text{MSE}_{c}$
If $\text{MSE}_1 + \text{MSE}_2$ equals $\text{MSE}_{c}$, then the MSE respects the Additivity axiom, that is not always true.

\newpage
\section{}\label{secA2}
\subsection{Forecast biases for years between 2000--2009}
\begin{figure}[!htbp]
     \centering
     \subfloat[][Female data]{\includegraphics[width = 0.95\linewidth]{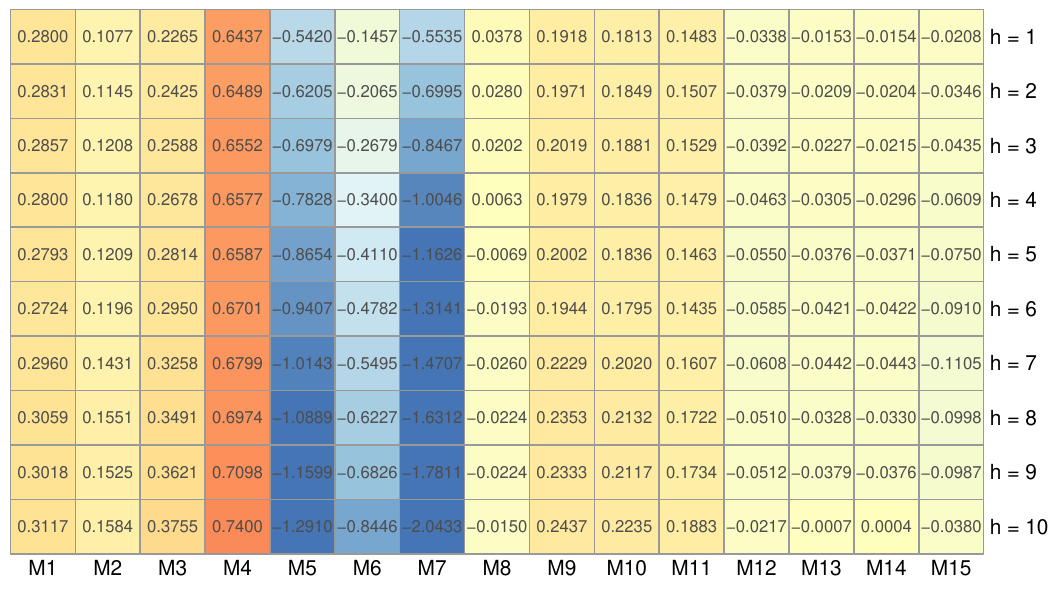}\label{Female}} \\
     \subfloat[][Male data]{\includegraphics[width = 0.95\linewidth]{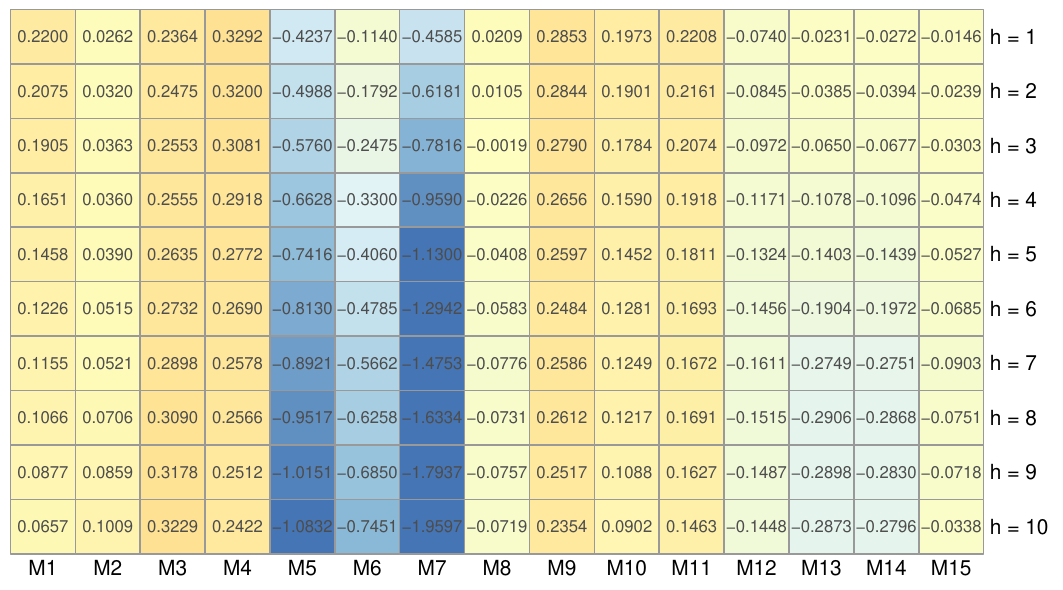}\label{Male}}
     \caption{Forecast biases produced by 15 individual mortality forecasting models for years between 2000 and 2009.}
\end{figure}\label{Biases}

\newpage
\subsection{Correlations of forecasts for years between 2000--2009}
\begin{figure}[!htbp]
     \centering
     \subfloat[][Female data]{\includegraphics[width = 0.95\linewidth]{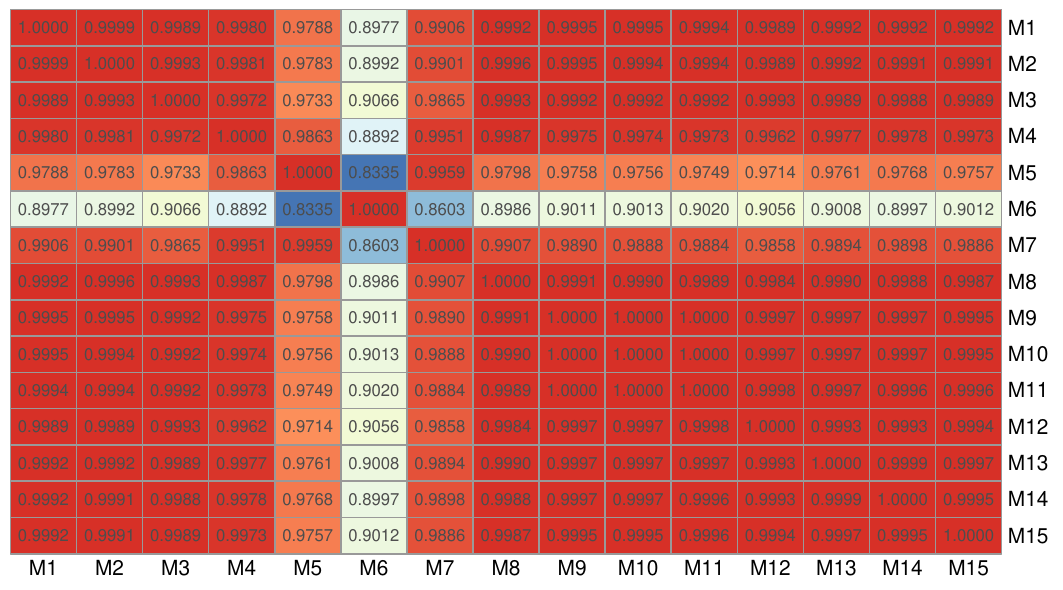}\label{Female_2}} \\
     \subfloat[][Male data]{\includegraphics[width = 0.95\linewidth]{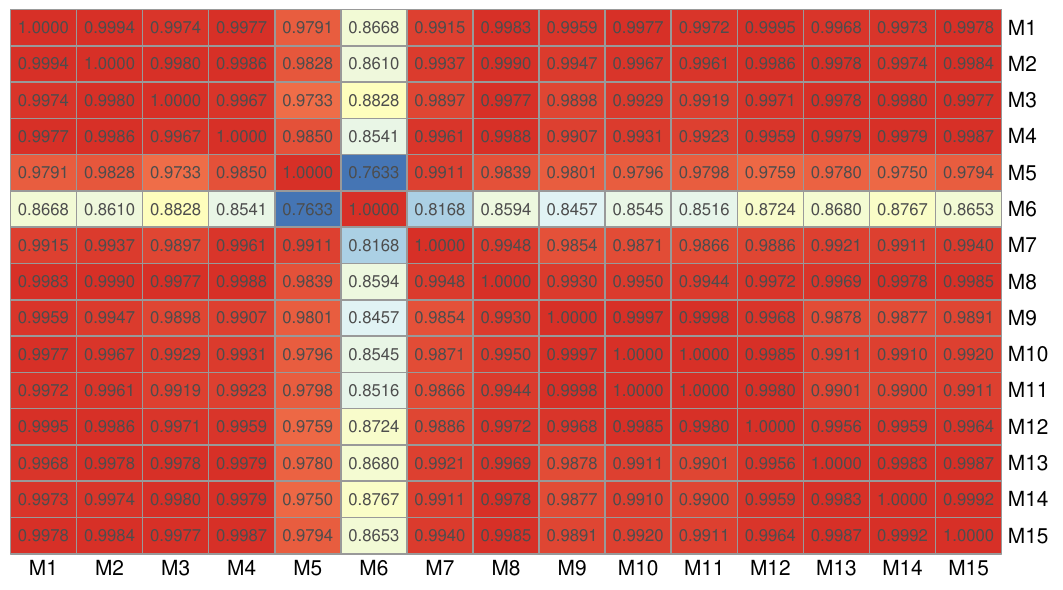}\label{Male_2}}
     \caption{Correlations of forecasts produced by 15 individual mortality forecasting models for years between 2000 and 2009.}
\end{figure}\label{Correlation}
\end{appendices}

\newpage
\bibliographystyle{agsm}
\bibliography{Shapley_value}

\end{document}